\documentclass[doublecol]{epl2} 
\usepackage{psfrag}
\usepackage{epsfig}

\newcommand     {\beq}[1]         { \begin{equation} #1 \end{equation} }
\newcommand     {\beqa}[1]        { \begin{eqnarray} #1 \end{eqnarray} }

\newcommand     {\ep}             { \varepsilon }
\newcommand     {\si}             { \sigma }

\title{Slip avalanches in a fiber bundle model}
\shorttitle{Slip avalanches in a fiber bundle model}

\author{Zolt\'an Hal\'asz \and Ferenc Kun
}
\shortauthor{Z. Hal\'asz and F. Kun}

\institute{                    
  \inst{} Department of Theoretical Physics, University of Debrecen \\
            - H-4010 Debrecen, P.O.Box: 5, Hungary\\
}

\abstract{
We study slip avalanches in disordered materials under an increasing
external load in the framework of a fiber bundle model. Over-stressed
fibers of the model do not break, instead they relax in a stick-slip
event which may trigger 
an entire slip avalanche. Slip avalanches are characterized by the
number slipping fibers, by the slip length, and by the load increment,
which triggers the avalanche.  Our calculations
revealed that all three quantities are characterized by power law
distributions with universal exponents. We show by analytical
calculations and computer simulations that varying the amount of
disorder of slip thresholds and  
the number of allowed slips of fibers, the system exhibits a disorder
induced phase transition from a phase where only small avalanches are
formed to another one where a macroscopic slip appeares. 
} 
\begin{document}

\maketitle

There is a large variety of non-equilibrium systems which exhibit
crackling noise, i.e.\ they respond to a slow continuous external
driving in the form of bursts of local events
\cite{sethna_nature_2001}. Examples can be
mentioned from earthquakes and fracture of disordered materials
\cite{zapperi_alava_statmodfrac,eq_davidsen_prl98_2007}, 
through Barkhausen noise in 
ferromagnets \cite{durin_prl2000}, to martensitic shape memory alloys and plastic
deformation of solids
\cite{miguel_nature_2001,zanzotto_prl2009}. During the last decade
experimental and theoretical investigations revealed that the
probability distributions of the characteristic quantities of bursts
have scale free behavior with universal exponents \cite{sethna_nature_2001,zapperi_alava_statmodfrac,zanzotto_prl2009,damen_zion_prl2009}. 
An intense research has been initiated to understand
the underlying mechanism of the observed universality. The
investigation of simple models which grasp the crucial features of
systems exhibiting crackling noise proved to be essential.
Along this line, based on the analogy of 
the plastic deformation and fracture of
heterogeneous materials, recently, a micromechanical model was
introduced in Ref.\ \cite{damen_zion_prl2009} which can reproduce the
main features of crackling noise in these types of systems with only
one tuning parameter.  

In the present Letter we study the emergence of crackling noise in
heterogeneous materials which respond to an increasing
external load by local rearrangements with stick-slip mechanism. We
consider a fiber bundle model
\cite{kloster_pre_1997,hansen_crossover_prl,kun_basquin_prl2008,naoki_prl2008,kovacs_mix_gls}
where over-stressed fibers do not break, instead they increase 
their relaxed length in a slip event until they can sustain the load.
The system is driven by small load increments giving rise to the
slip of a single fiber which may then trigger an entire avalanche of
slip events due to load redistribution in the bundle. We show by
analytic calculations and computer simulations that the load increment
triggering the slip bursts, furthermore, the number of slipping fibers
and the total slip length of the bundle are all characterized by power law
distributions. We demonstrate that the amount of disorder and the
total number of allowed slips play a crucial role in the system: a
disorder induced phase transition
\cite{sethna_nature_2001,zanzotto_prl2009,dahmen_prl1993,sornette_prl_78_2140}
is obtained from a low disorder phase 
where the system snaps with macroscopic bursts to the high disorder
one where only small avalanches pop up. Our model provides an adequate
description of the micromechanics of disordered systems which 
store hidden length
\cite{halasz_kun_1}, and it can also be considered as the fiber bundle
analogue of the Burridge-Knopoff model of earthquakes with an infinite
range of interaction \cite{burridge_knopoff}.

Our model consists of $N$ fibers assembled in parallel. Under an
increasing external load $\si$ the fibers exhibit a linearly elastic
behavior characterized by the same Young modulus $E$. The important
novel element of the model is that when the deformation $\ep$ of a
fiber reaches a threshold value $\ep_{th}$ the fiber does not
break. Instead, its relaxed length increases until the load reduces to
zero on the fiber. The mechanism of relaxation is the slip of the
fiber end, or it can also be interpreted as the unfolding of subunits
of fibers which provide some stored length \cite{halasz_kun_1}. The
slip thresholds 
$\ep_{th}^i$, $i=1,\ldots , N$ are random variables with a probability
density $p(\ep_{th})$ and distribution function $P(\ep_{th})$.
After
the slip event the fiber gets sticked again so that it can support
load and can suffer further slips. The load kept by fiber $i$ at a
deformation $\ep$ after slipping reads as $\si_i=E(\ep-\ep_{th}^i)$ so
that no hardening or softening is assumed in the system. When a fiber
slips again either the same slip threshold is retained (quenched
disorder) or new threshold values can be drawn from the same
probability distribution $p(\ep_{th})$ (annealed disorder). 
The total number of
slip events  $k_{max}$  a fiber can suffer is a very important parameter
of the model which can vary in the range $1\leq k_{max}< +\infty$. 
For the load redistribution following slip events we assume an
infinite range of interaction, i.e.\ equal load sharing, which is
ensured by the condition that the strain $\ep$ is the same for all
fibers. In the present paper our 
analysis is restricted to the case of quenched disorder so that the
load of fiber $i$ in the bundle after $k_i\leq k_{max}$ slips takes the
form  $\si_i=E(\ep-k_i\ep_{th}^i)$. 
Further details of the model construction
can be found in Ref.\ \cite{halasz_kun_1} including also the case of
annealed slip thresholds. 

Based on the assumption of equal load sharing the
constitutive equation of the 
parallel bundle can be obtained analytically 
\begin{figure}
  \begin{center}
  \epsfig{bbllx=15,bblly=500,bburx=325,bbury=770,
file=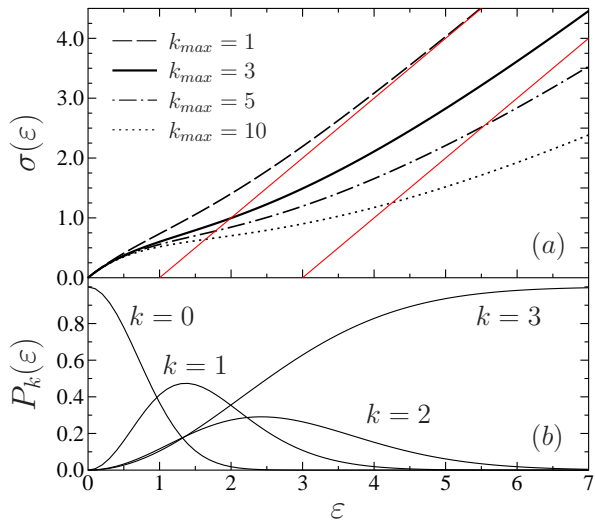,width=8cm}
  \caption{{\it (Color online)} $(a)$ Constitutive behavior of the
bundle with exponentially 
distributed quenched failure thresholds. $\sigma(\ep)$ tends to an
asymptotic linear response preceded by a longer horizontal
plateau when $k_{max}$ increases. The intersection of the asymptotic
straight lines with the horizontal axis indicate the remaining
deformation when unloading the system. $(b)$ Breaking probabilities
$P_k(\varepsilon)$ for $k_{max}=3$.
}
  \label{fig:const_exp}
  \end{center}
\end{figure}
by integrating 
the load kept by the subsets of fibers with different slip indices
$k$ \cite{halasz_kun_1}
\beqa{
&&  \sigma(\ep) =  E\ep\left[1-P(E\ep) \right]  \nonumber  \\
&&+
\displaystyle{\sum_{k=1}^{k_{max}-1}\int_{\ep/(k+1)}^{\ep/k}p(E\ep_1)
E\left(\ep-k\ep_1\right)d\ep_1}  \nonumber \\ 
&&+  \displaystyle{\int_{0}^{\ep/k_{max}}p(E\ep_1)E
\left(\ep-k_{max}\ep_1\right)d\ep_1}.  
\label{eq:constit_3}}
Note that the integrals have to be performed over the entire loading
history of the bundle.  
For very large deformations $\ep \to
\infty$, practically all fibers have suffered $k_{max}$ slips
so that Eq.\ (\ref{eq:constit_3}) can be rewritten as
$\sigma(\ep) \sim E\ep -k_{max}E \int_{0}^{\ep/k_{max}}p(\ep_1)\ep_1 d\ep_1$
where the integral provides the average value of the 
slip thresholds $\left<\ep_{th} \right>$.
It means that the bundle has an asymptotic linear behavior with the
initial value of the Young modulus, however, when unloading the
system $\sigma \to 0$ an irreversible permanent deformation
remains whose maximum $\varepsilon_r^{max}$ value is proportional to
the average slip length 
$\left<\ep_{th}\right>$ and the number of slip events $k_{max}$
allowed $\ep_r^{max} = k_{max}\left<\ep_{th}\right>$. 
We note that for a finite bundle of $N$ fibers with quenched slip
thresholds the constitutive equation Eq.\ (\ref{eq:constit_3}) can be
written in a discrete form
\begin{eqnarray}
\sigma(\varepsilon)=E\varepsilon-(E/N)
\sum_{i=1}^{N}k_i(\varepsilon)\varepsilon_{th}^i,
\label{eq:discrete}
\end{eqnarray} 
where $k_i(\varepsilon)$ denotes the number of slips suffered by fiber
$i$  up to deformation $\varepsilon$.

In the explicit calculations we use Weibull distributed threshold
values $\ep_{th}$ with the probability density function
\beq{
\displaystyle{p(\ep_{th}) =
m\frac{\ep_{th}^{m-1}}{\lambda^{m-1}}e^{-\left(\ep_{th}/
\lambda\right)^m}}, 
\label{eq:weibull}
}
where the parameter $\lambda$ setting the scale of the thresholds is
fixed to $\lambda = 1$ in our entire study. The Weibull exponent
$1\leq m < +\infty$ is a
very important characteristics of the system, which controls the amount
of disorder in the slip thresholds. Increasing the value of $m$
from 1 to infinity the probability density Eq.\ $(\ref{eq:weibull})$
varies from the exponential distribution to the delta function
of zero width. Figure \ref{fig:const_exp} illustrates the constitutive
curve $\sigma(\ep)$ of the 
model with exponentially distributed quenched slip thresholds ($m=1$) for several 
different values of $k_{max}$. It can be seen that increasing the
maximum number of breakings allowed a plastic plateau develops, i.e.\
the final asymptotic linear part 
of the constitutive curve is preceded by a longer and longer
horizontal plateau. The slope of $\sigma(\ep)$ in the asymptotic
regime is equal to the Young modulus $E=1$ of fibers. Note that the
simple form of $\sigma(\varepsilon)$ in Fig.\ \ref{fig:const_exp}$(a)$
is the consequence of the monotonous behavior of the exponential
distribution, i.e. varying the value of $m$ and $k_{max}$ along the
plastic plateau $\sigma(\varepsilon)$ can have a more complex
functional form which will be explored below.
Macroscopic failure of the system can be
captured in the model by assuming that the fibers break after having
suffered $k_{max}$ slips, which has been studied in Ref.\
\cite{halasz_kun_1}. In the present 
paper we focus on the microscopic stick-slip process of the fiber
bundle with quenched slip thresholds retaining the fibers' stiffness
after $k_{max}$ slips (no breaking).

Quasi-static stress controlled loading of the fiber bundle can be
performed by
incrementing the external load with a small amount $\delta\si$ just to
provoke the slip of a single fiber. 
Since the external load $\si$ is kept constant during the slip, the
load dropped by the slipping fiber must 
be overtaken by the other ones which can give rise to further slip
events. This way a single slip induced by the load increment
$\delta\si$, can trigger an entire avalanche of slips, which increases
the macroscopic strain $\ep$ of the system by the amount $\delta \ep$.
This jerky microscopic dynamics has
the consequence that the deformation of the bundle has a step-wise
increase under a quasi-statically increasing external load $\sigma$.
We characterize the slip avalanches by their size $\Delta$
defined as the number of fibers slipping in the avalanche, and by the
emerging slip length $\delta \ep$ which is the increment of the strain
$\ep$ of the bundle. All the three quantities, the load increment
$\delta\si$ which triggers the avalanche, the avalanche size $\Delta$,
and the slip length 
$\delta\ep$ are random variables so that the stick-slip process on the
micro-scale can be characterized by their probability distributions
$P(\delta\si)$ $P(\Delta)$, and $P(\delta \ep)$, respectively. Note
that under strain controlled loading no slip avalanches can arise. 

For simple fiber bundles where fibers break irreversibly when the
local load surpasses their threshold value, it has
recently been shown \cite{kloster_pre_1997,hansen_crossover_prl} for
the case of equal load sharing that the size 
distribution of avalanches 
$P(\Delta)$ can be obtained in a closed analytical form as
\beq{
P(\Delta) \approx \frac{e^{\Delta}}{\sqrt{2\pi}\Delta^{3/2}}
\int_0^{\ep_c} p(\ep)\frac{1-a(\ep)}{a(\ep)}
e^{\Delta\left[a(\ep)-\ln{a(\ep)}\right]}d\ep.  
\label{eq:hansen}
}
Here $a(\ep)$ denotes the average number of fibers which break as a
consequence of a single fiber failure induced by the external load
increment at the deformation $\ep$. The integration over $\ep$ is carried
out up to the critical point $\ep_c$ of the system where catastrophic
collapse occurs. The dominating contribution to the integral is
provided by the vicinity of the maximum of the exponent of the
integrand $\psi(\ep)=a(\ep)-\ln{a(\ep)}$, which is obtained
at $a=1$. After 
Taylor expansion of $a(\ep)$ and of the exponent $\psi(\ep)$ about the
maximum the asymptotics of the size distribution $P(\Delta)$ reduces
to the power law form 
\beq{
P(\Delta) \sim \Delta^{-\tau}.
}
The exponent $\tau=5/2$ proved to be universal for a broad class
of disorder distributions where the macroscopic constitutive curve
$\sigma(\ep)$ of the system has a single quadratic maximum
\cite{kloster_pre_1997,hansen_crossover_prl,kovacs_mix_gls}.

In order to understand the dynamics of slip avalanches in our model,
first the sequence of slipping events has to be analyzed. 
The probability $P_k(\ep)$ that a randomly selected fiber in the bundle
has suffered exactly $k$ slips up to the deformation $\ep$ can be
obtained analytically as 
\beqa{
P_0(\ep) &=& 1-P(\ep), \\
P_k(\ep) &=& P\left(\frac{E\ep}{k}\right) -
P\left(\frac{E\ep}{k+1}\right), \ \ \ 1\leq k < 
k_{max}, \nonumber \\
P_{k_{max}} &=& P\left(\frac{E\ep}{k_{max}}\right), \nonumber
\label{eq:big_p}
}
where $P$ denotes the cumulative distribution of the slip
thresholds. The functional form of $P_k(\ep)$ is presented in Fig.\
\ref{fig:const_exp}$(b)$ for $m=1$ and $k_{max}=3$. 
From the above equations one can determine the probability
density $p_k^{k+1}(\ep)$ of events 
that a fiber which has suffered $k$ slips until the
 deformation $\ep$ was reached, will slip again due to the strain
increment $d\ep$
\beqa{
p_k^{k+1}(\ep) &=& \frac{1}{k+1}p\left(\frac{\ep}{k+1}\right), \ \ \
0\leq k< k_{max},
\label{eq:small_p}
}
where $p$ is the original probability density of the slip
thresholds. When the external load is increased by the amount 
$\delta\si$ at the
deformation $\ep$ to provoke the slip of a single fiber $i$ which has
already slipped $k$ times, the strain increment $\delta
\ep=\delta\ep_k$ arising due to load 
redistribution after the slip can be obtained from the constitutive
equation of finite bundles Eq.\ (\ref{eq:discrete}).
Keeping the load 
$\si$ fixed during the slip, the strain increment reads as $\delta
\ep_k=\ep^{i}_{th}/N=\ep/(kN)$. It follows that
the average number of fibers $a(\ep)$ which slip as a consequence of
a single slip can be  determined as $a(\ep)=N\sum_{k=0}^{k_{max}-1}\delta
\ep_kp_k^{k+1}(\ep)$ which leads to the form
\beq{
a(\ep) = \ep\sum_{k=1}^{k_{max}}\frac{1}{k^2}
p\left(\frac{E\ep}{k}\right).  
\label{eq:af}
}

 \begin{figure}
  \begin{center}
\epsfig{bbllx=0,bblly=450,bburx=400,bbury=770,
file=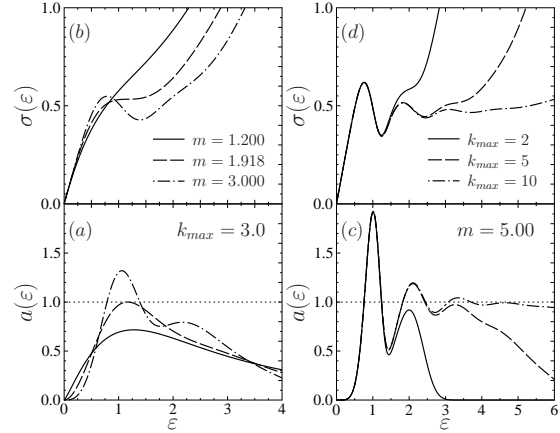,width=7.3cm}
   \caption{Constitutive curves $\sigma(\ep)$ and average number of
induced slips $a(\ep)$ for a fixed $k_{max}=3$ varying the value of
$m$ $(a,b)$, and for a fixed $m=5.0$ varying the value of $k_{max}$ $(c,d)$. 
Note that $m^c_3\approx1.918$. } 
\label{fig:constit_compare}
  \end{center}
\end{figure}
It is important to emphasize that the derivative of the constitutive
equation $\sigma(\ep)$ of Eq.\ 
(\ref{eq:constit_3}) can be expressed in terms of $a(\ep)$ as
\beq{
\frac{d\sigma}{d\ep} = E\left[1-\sum_{k=1}^{k_{max}}
\frac{\ep}{k^2}p\left(\frac{E\ep}{k}\right)\right]=E\left[1-a(\ep)\right],
\label{eq:sigma_deriv}
}
which show that the constitutive curve $\sigma(\ep)$ has extrema at
locations $\ep_c$ where
the average number of induced slips becomes unity $a(\ep_c) = 1$. It
also follows from Eq.\ (\ref{eq:sigma_deriv}) that at the extremal
points of $a(\ep)$ the constitutive curve $\sigma(\varepsilon)$ has an
inflexion point $d^2\sigma/d\ep^2=0$.

Our calculations revealed that varying the amount of
disorder $m$ and the number of allowed slip events $k_{max}$ the
statistics of avalanches exhibits a very complex behavior. Starting
from Eqs.\ 
(\ref{eq:constit_3},\ref{eq:af},\ref{eq:sigma_deriv}) we can
determine analytically the phase diagram of the system on 
the $m-k_{max}$ plane which classifies all possible functional forms
of the constitutive curves $\si(\ep)$ and of avalanche size
distributions $P(\Delta)$.  
Writing $a(\ep)$ in the form $a(\ep) = \sum_{k=1}^{k_{max}} a_k(\ep)$, 
it can be seen that each term $a_k(\ep) = (\ep/k^2)p(E\ep/k)$ has a
single maximum at the strain $\ep_k^c$ where $\ep_k^cp(E\ep_k^c/k)=-1$
holds. It follows that if $a_1$ has a maximum
at $\ep_1^c$ with the value $a_1^c$, then the maxima of the other
terms $a_k(\ep)$ are placed equidistantly as $\ep_k^c = k\ep_1^c$ with
decreasing 
values $a_k^c=a_1^c/k$. Due to the overlap of the functions
$a_k(\ep)$, the consecutive maxima of $a(\ep)$ do not coincide with
that of $a_k(\ep)$, however, the equidistant spacing and the
decreasing sequence survive. For the case of Weibull distributions the
above analysis results in $\ep_k^c=k\lambda$ and $a_k^c=m/(ke)$, where
$e$ is the base of natural logarithm. It can be seen that for $k_{max}=1$,
when only a single slip is allowed, at the critical Weibull exponent $m_1^c=e$ the
constitutive curve $\sigma(\ep)$ has an inflexion point at the
position $\ep_{1}^c$ where
$a(\ep)$ has a maximum with the value $a(\ep_1^c)=1$. Similarly, for any
$k_{max}\geq 1$ one can find an 
$m_{k_{max}}^c$ value of the Weibull exponent, where the constitutive
curve has an inflexion point with the properties
$d\si/d\ep|\ep_{k_{max}}^c=0$ and $d^2\si/d\ep^2|\ep_{k_{max}}^c=0$, where
at the same time $a(\ep_{k_{max}}^c)=1$ and
$da/d\ep|\ep_{k_{max}}^c=0$ hold.

\begin{figure}
  \begin{center}
\epsfig{bbllx=40,bblly=20,bburx=430,bbury=370,
file=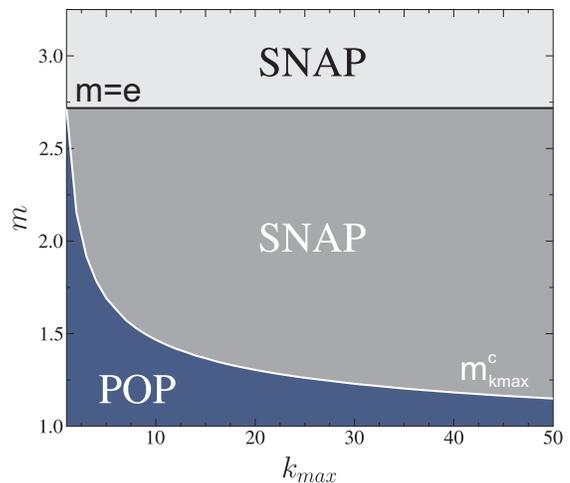,width=7.3cm}
   \caption{{\it (Color online)} Phase diagram of the system. The decreasing line indicates
the $m_{k_{max}}^c$ curve which separates the $POP$ and $SNAP$
regimes. 
} 
\label{fig:phase_diag}
  \end{center}
\end{figure}

The phase diagram of the system is presented in Fig.\
\ref{fig:phase_diag}, where the decreasing line represents the
$m_{k_{max}}^c$ curve which was determined numerically. Note that for
$k_{max}=1$ the critical Weibull exponent is 
$m_1^c=e$ and $m_{k_{max}}^c\to 1$ holds for $k_{max}\to +\infty$. 
In order to obtain the asymptotics of the size distribution of slip
avalanches $P(\Delta)$ analytically from Eq.\ (\ref{eq:hansen}) for
parameters along the $m^c_{k_{max}}$ curve,
the Taylor expansion of $a(\ep)$ and $\psi(\ep)$ about
$\varepsilon^c_{k_{max}}$ has to be continued 
beyond the first order terms. Following the derivations of Ref.\
\cite{kovacs_mix_gls} the  first non-vanishing terms are $a(\ep) \simeq
1+C_1(\ep-\ep_{k_{max}}^c)^2$ and $\psi(\ep)\simeq
1+C_2(\ep-\ep_{k_{max}}^c)^4$, which result in a power 
law asymptotics $P(\Delta)\sim \Delta^{-\tau}$ with the exponent
$\tau=9/4$. A similar behavior was found in Ref.\
\cite{kovacs_mix_gls}, where a 
different physical mechanism led to a similar constitutive curve of
the system. 

\begin{figure}
  \begin{center}
\epsfig{bbllx=20,bblly=380,bburx=600,bbury=650,
file=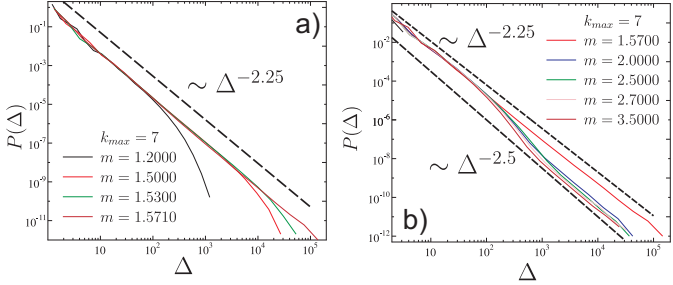,width=8.8cm} 
   \caption{{\it (Color online)} $(a)$ Size distribution of slip
     avalanches $P(\Delta)$ in the  
$POP$ phase for $k_{max}=7$ varying the amount of disorder $m$. A high
quality power law behaviour is obtained with a diverging cutoff as the
phase boundary is approached $m\to m^c_7\approx 1.572$. $(b)$
Comparison of avalanche size distributions in the $POP$ and $SNAP$
phases. Excellent agreement is obtained with the analytic predictions.
}  
\label{fig:scaled_1}
  \end{center}
\end{figure}

The parameter regime of $m$ and $k_{max}$ below the $m^c_{k_{max}}$
curve of Fig.\ \ref{fig:phase_diag} defines the high disorder phase of the
model, where $\si(\ep)$ is 
monotonically increasing $d\si/d\ep>0$ and the maximum of $a(\ep)$ is
always smaller than 1. Figure 
\ref{fig:constit_compare}$(a,b)$ illustrates the constitutive behavior
$\si(\ep)$ and the 
average number of induced slips $a(\ep)$ for $k_{max}=3$ varying the
value of $m$, where the
critical disorder parameter is $m^c_{3}\approx 1.918$. 
Since the minimum value of the derivative $d\sigma/d\varepsilon$ Eq.\
(\ref{eq:sigma_deriv}) is
positive in the high disorder phase, the avalanche size distribution
$P(\Delta)$ behaves as in simple fiber bundles 
when the loading process was stoped at a deformation $\varepsilon_m$
before the critical point of macroscopic failure
\cite{kloster_pre_1997}. Hence, for $m<m^c_{k_{max}}$ from Eq.\ 
(\ref{eq:hansen}) the size distribution of bursts takes the form 
\begin{eqnarray}
P(\Delta) \sim \Delta^{-\tau}e^{-\left[a(\varepsilon_m)-1-\ln
a(\varepsilon_m)\right]\Delta}, 
\label{eq:cutoff}
\end{eqnarray}
i.e.\ the power law regime of exponent $\tau=9/4$ is followed by an
exponential cutoff, where in our case $\varepsilon_m$ is the position
of the inflexion point of the constitutive curve.
Since in the high disorder phase of
the model only relatively small avalanches pop up away from the phase
boundary, we call this phase 
as $POP$ phase, following the terminology of Ref.\
\cite{zanzotto_prl2009}. Figure
\ref{fig:scaled_1}$(a)$ presents the size distribution of slip avalanches
$P(\Delta)$ obtained by 
computer simulations for $k_{max}=7$ at different $m$ values in the
range $m\leq m^c_7$. A high quality power law behavior can be
observed with a diverging cutoff as approaching the
critical point $m\to m^c_7$ in agreement with the above derivation. 

In the low disorder regime, above the $m^c_{k_{max}}$ curve of Fig.\
\ref{fig:phase_diag}, the 
constitutive curve can have local maxima along the plateau regime. It
follows from the above derivation that at a given value of $k_{max}$
the number of maxima of $\sigma(\varepsilon)$ is one if the value of
$m$ falls in the interval $m^c_{k_{max}}<m<e$. Under stress controlled
loading a macroscopic avalanche appears resulting in a horizontal
jump when the maximum of $\sigma(\varepsilon)$ is reached.
Consequently, this phase of the model is called $SNAP$ phase
\cite{zanzotto_prl2009}.  
At very low disorder $m>e$ the constitutive curve has a local maximum
already at $k_{max}=1$ and further maxima occur with decreasing height
accompanied by a similar oscillating behavior of
$a(\varepsilon)$ as $k_{max}$ increases. 
This feature can be observed in Figures
\ref{fig:constit_compare}$(c,d)$, which present the behavior of 
$\sigma(\varepsilon)$ and $a(\varepsilon)$ for $m=5$ varying the value
of $k_{max}$. 
In the $SNAP$ phase the distribution of avalanche sizes
$P(\Delta)$ is determined by the first maximum of $\si(\ep)$ which has
a quadratic shape. Consequently, similarly to the case of simple fiber
bundles, $P(\Delta)$ has a power law
functional form $P(\Delta)\sim \Delta^{-\tau}$ without cutoff regime
but with an exponent 
$\tau=5/2$ higher than in the $POP$ phase
\cite{kloster_pre_1997,hansen_crossover_prl,kovacs_mix_gls}. Burst
size  distributions of the $POP$ and $SNAP$ phases are compared in
Fig.\ \ref{fig:scaled_1}$(b)$, where nice agreement can be observed with the 
analytic predictions.

\begin{figure}
  \begin{center}
\epsfig{bbllx=5,bblly=450,bburx=390,bbury=790,
file=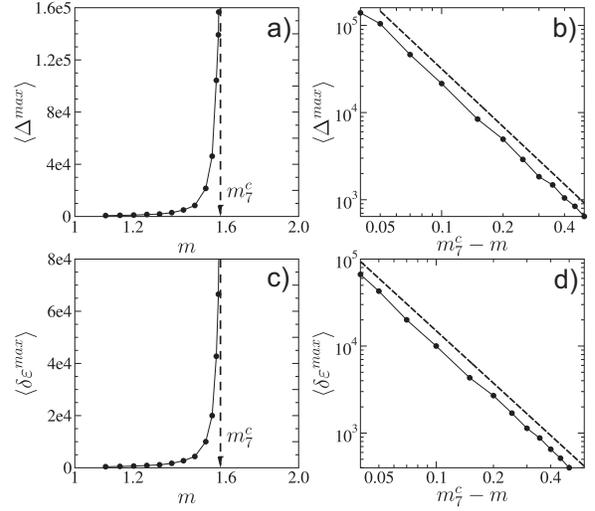,width=7.7cm} 
   \caption{The average size of the largest avalanche
$\left<\Delta^{max}\right>$ $(a)$ and 
the average largest slip length
$\left<\delta\ep^{max}\right>$ $(c)$ as a function of the Weibull
exponent $m$ below the critical 
point $m_7^c$. The same quantities are presented in $(b,d)$ as a
function of the distance from the critical point $(m_7^c-m)$.
 } 
\label{fig:delta_max}
  \end{center}
\end{figure}

In order to understand the transition from the $POP$ to the $SNAP$
phase when the amount of disorder $m$ is varied, we further analyze the
burst size distribution Eq.\ (\ref{eq:cutoff}). For the specific case
of $k_{max}=1$ we
have $a(\varepsilon_m)=a^c_1$, where $a^c_1=m/e$ converging to 1 when
the amount of disorder is 
decreased in the $POP$ phase $m\to m^c_{1}=e$. After
Taylor expanding the terms in the exponential function of Eq.\
(\ref{eq:cutoff}) about $m^c_1$, we obtain the form 
\begin{eqnarray}
P(\Delta) \sim \Delta^{-\tau}e^{-\Delta/\Delta_0},
\label{eq:pop_dist}
\end{eqnarray} 
where the characteristic burst size has a power law divergence
\begin{eqnarray}
\Delta_0\sim (m^c_{k_{max}}-m)^{-\nu}
\label{eq:pop_diverge}
\end{eqnarray}
with the cutoff exponent $\nu=2$. The exact proof is for $k_{max}=1$
but our numerical calculations revealed that the analytic results
Eqs.\ (\ref{eq:pop_dist},\ref{eq:pop_diverge}) hold for all
values of $k_{max}$ in the $POP$ phase. In order to numerically verify
the above analytic predictions, we assume that the cutoff avalanche
size $\Delta_0$ is proportional to the average size of the largest
avalanche $\Delta_0\sim \left<\Delta^{max}\right>$.
Figure \ref{fig:delta_max}$(a)$ presents $\left<\Delta^{max}\right>$
obtained by computer 
simulations of 
a bundle of $N=10^7$ fibers with $k_{max}=7$ varying the Weibull 
exponent $m$ in a broad range. It can be seen that approaching the phase
boundary $m^c_{k_{max}}$ from the $POP$ phase $\left<\Delta^{max}\right>$
diverges, i.e.\ it exhibits a sharp maximum in the finite
system. Figure \ref{fig:delta_max}$(b)$ presents the same data as a
function of the distance from the critical point $m_7^c$, where a power law
behavior is evidenced with an exponent $\nu=2.05\pm 0.05$ in a good
agreement with Eq.\ (\ref{eq:pop_diverge}).
The results imply that
varying the amount of threshold disorder the bundle of stick-slip
fibers undergoes a disorder induced phase transition from the high
disorder phase where small avalanches pop up to the low disorder one
where macroscopic avalanches snap the system 
\cite{sethna_nature_2001,zanzotto_prl2009,dahmen_prl1993}.  

\begin{figure}
  \begin{center}
\epsfig{bbllx=5,bblly=465,bburx=400,bbury=750,
file=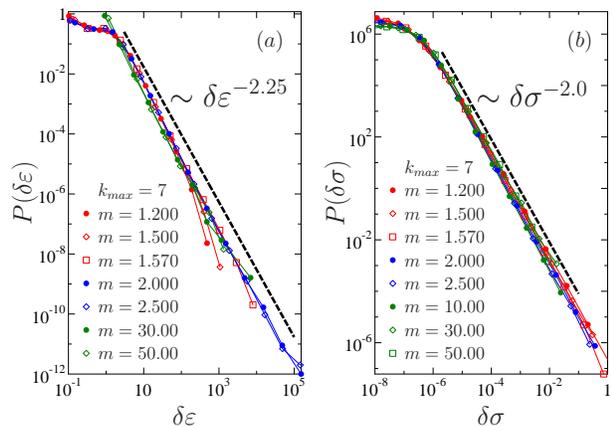,width=8.0cm} 
   \caption{{\it (Color online)} Probability distribution of slip
length $P(\delta\ep)$ of avalanches $(a)$ and the distribution of load
increments $P(\delta\si)$ which trigger the avalanches  $(b)$ for different
values of the Weibull exponent $m$. Universal power law behavior is
obtained in both cases.}  
\label{fig:dsigma}
  \end{center}
\end{figure}

In spite of the complexity of the behavior of avalanche sizes,
computer simulations revealed a universal
functional form for both the distribution of  the slip length
$P(\delta\ep)$ and for the load increments $P(\delta\sigma)$. 
The total slip length, i.e.\ the strain increment $\delta\ep$ occurred
during an avalanche of size $\Delta$ reads as $\delta \ep =
(1/N)(\ep_{th}^{i_1}+\ep_{th}^{i_2}+\cdot
+ \ep_{th}^{i_{\Delta}})$, where $\ep_{th}^{i_j}$ denotes the slip
threshold of fibers taking part in the avalanche. The distribution of
slip length $P(\delta\ep)$ is presented in Fig.\ \ref{fig:dsigma}$(a)$
for several values of $m$. 
It can be observed that $P(\delta\ep)$
exhibits a universal power law behavior 
\beq{
P(\delta\ep) \sim \delta\ep^{-\phi},
}
where the value of the exponent $\phi=2.25\pm 0.05$ was obtained
numerically independently of $m$ and $k_{max}$. Similarly to the
avalanche size $\Delta$, the characteristic slip length defined as the
average value of the largest slip length
$\left<\delta\ep^{max}\right>$ is sensitive to the precise shape
of $\si(\ep)$ in the vicinity of the extremal points. It can be
observed 
in Fig.\ \ref{fig:delta_max}$(c)$ that
$\left<\delta\ep^{max}\right>$ has a sharp 
maximum approaching the phase boundary from the $POP$ phase 
and a power law divergence of the type of Eq.\ (\ref{eq:pop_diverge})
is evidenced in Fig.\ \ref{fig:delta_max}$(d)$. The critical
exponent $\nu\approx 2$  proved to be the same as for the
characteristic burst size. 
Figure \ref{fig:dsigma}$(b)$ presents that the distribution of load
increments has a power law decay
\beq{
P(\delta\si) \sim \delta\si^{-\alpha},
}
where the value of the exponent is universal $\alpha=2$, it does not
depend neither on $m$ nor on $k_{max}$. It has to be emphasized that
the asymptotics of the distribution $P(\delta\si)$ is determined by
the beginning of the loading process where large load increments
are required to drive the system. The reason of
universality is that the low stress regime of the loaded system is
insensitive to the parameters $m$ and $k_{max}$.

In summary, we studied the statistics of slip avalanches in a fiber
bundle model 
where overstressed fibers can relax in a series of stick slip
events. We showed that on the macro-scale the stick-slip mechanism
leads to plastic behavior with a permanent deformation remaining
after the load is released. On the micro-scale single slips induced
by external load increments trigger bursts which give rise to a
step-wise strain increase. 
The distribution of load increments and of slip length exhibit
a universal power law behavior with exponents independent of the
model's parameters. The size distribution of bursts proved to be
sensitive to the amount of disorder and to the number of fibers'
slips. Our calculations revealed that at high enough disorder only
small avalanches pop up, while at low disorder macroscopic
avalanches can snap the system. We set up the phase diagram of the
model and showed that the 
transition between the $POP$ and $SNAP$ phases is analogous to disorder
induced phase transitions. Besides the theoretical interest, our
calculations provide insight into the statistics of restructurings of
systems with hidden length such as biomaterials. Our study was
restricted to the case of Weibull distributions where the amount of
disorder can be represented by the exponent $m$. Generalization to
other distributions defined over an infinite domain is straightforward
using the standard deviation as a measure of disorder. The value of
the critical exponents $\tau, \nu, \phi$, and $\alpha$ do not have any
dependence on the functional form of the disorder distribution.

It has been shown in Ref.\ \cite{zanzotto_prl2009} that the mode of
external driving has a crutial effect on the critical non-equilibrium
steady states in slowly driven bistable heterogeneous systems with
controllable disorder: changing the driving from soft to hard a
crossover is obtained from the classical order-disorder universality
class to the quenched Edwards-Wilkinson class of SOC type. I our
investigations of the avalanche statistics only stress controlled
loading was considered, which corresponds to the perfectly soft
driving of Ref.\ \cite{zanzotto_prl2009}. Our phase diagram of Fig.\
\ref{fig:phase_diag} can be considered as an extension of the zero
softness part of the phase diagram of Ref.\ \cite{zanzotto_prl2009}
with the additional degree of freedom of varying the number of allowed
slip events under an infinite range of interaction. It is very
interesting to extend our study to vary the mode of driving which is
currently in progress.

\acknowledgments
This work was supported by the MTA-JSPS program.
F.\ Kun acknowledges the Bolyai Janos fellowship of the Hungarian
Academy of Sciences.


\begin{thebibliography}{0}

\bibitem{sethna_nature_2001}
  \Name{Sethna J. P., Dahmen K. A., \and Meyers C. R.}
  \REVIEW{Nature}{410}{2001}{242}.

\bibitem{zapperi_alava_statmodfrac}
  \Name{Alava M., Nukala P. K., \and Zapperi S.}
  \REVIEW{Adv. Phys.}{55}{2006}{349}.

\bibitem{eq_davidsen_prl98_2007}
  \Name{Davidsen J., Stanchits S.  \and Dresen G.}
  \REVIEW{Phys. Rev. Lett.}{98}{2007}{125502}.

\bibitem{durin_prl2000}
  \Name{Durin G.  \and Zapperi S.}
  \REVIEW{Phys. Rev. Lett.}{84}{2000}{4705}.

\bibitem{miguel_nature_2001}
  \Name{Miguel M.-C., Vespignani A., Zapperi S., Weiss J.  \and Grasso J.-R.}
  \REVIEW{Nature}{410}{2001}{667}.

\bibitem{zanzotto_prl2009}
  \Name{P\'erez-Reche F.-J., Truskinovsky L., \and Zanzotto G.}
  \REVIEW{Phys. Rev. Lett.}{101}{2008}{230601}.

\bibitem{damen_zion_prl2009}
  \Name{Dahmen K. A., Ben-Zion Y., \and Uhl J. T.}
  \REVIEW{Phys. Rev. Lett.}{102}{2009}{175501}.

\bibitem{dahmen_prl1993}
  \Name{Sethna J. P., Dahmen K. A., Kartha S., Roberts B. W., \and
Shore J. L.} 
  \REVIEW{Phys. Rev. Lett.}{70}{1993}{3347}.

\bibitem{sornette_prl_78_2140}
  \Name{ Andersen J. V., Sornette D., \and Leung K.} 
  \REVIEW{Phys. Rev. Lett.}{78}{1997}{2140}.

\bibitem{kloster_pre_1997}
  \Name{Kloster M., Hansen A., \and Hemmer P. C.}
  \REVIEW{Phys. Rev. E}{56}{1997}{2615}.

\bibitem{hansen_crossover_prl}
  \Name{Pradhan S., Hansen A., \and Hemmer P. C.}
  \REVIEW{Phys. Rev. Lett.}{95}{2005}{125501}.

\bibitem{kun_basquin_prl2008}
  \Name{Kun F., Carmona H. A., Andrade Jr. J. S., \and Herrmann H. J.}
  \REVIEW{Phys. Rev. Lett.}{100}{2008}{094301}.

\bibitem{naoki_prl2008}
  \Name{Yoshioka N., Kun F., \and Ito N.}
  \REVIEW{Phys. Rev. Lett.}{101}{2008}{145502}.

\bibitem{kovacs_mix_gls}
  \Name{Hidalgo R. C., Kov\'acs K., Pagonabarraga I., \and Kun F.}
  \REVIEW{Europhys. Lett.}{81}{2008}{54005}.

\bibitem{halasz_kun_1}
  \Name{Hal\'asz Z., \and Kun F.}
  \REVIEW{Phys. Rev. E}{80}{027102}{2009}{}.

\bibitem{burridge_knopoff}
  \Name{Burridge R., \and Knopoff L.}
  \REVIEW{Bull. Seis. Soc. Amer.}{57}{1967}{341}.

\end{thebibliography}
\end{document}